\documentclass[12pt]{article}

\usepackage[a4paper,left=30mm,right=20mm,top=20mm,bottom=20mm]{geometry}
\usepackage{graphics}
\usepackage{amsmath}
\usepackage{epsfig}
\usepackage{cite}
\usepackage{xcolor}

\newcommand{\z}{&&\hspace*{-1cm}}

\newcommand{\bea}{\begin{eqnarray}}
\newcommand{\eea}{\end{eqnarray}}
\newcommand{\be}{\begin{equation}}
\newcommand{\ee}{\end{equation}}
\newcommand{\MSbar}{\overline{\rm MS}}

\newcommand{\Li}{\mathop{\mathrm{Li}}\nolimits}

\title{Transverse momentum dependent parton densities in
processes with heavy quark generations}

\author{A.V.~Kotikov$^{1}$, A.V.~Lipatov$^{1,2}$,
  P.~Zhang$^{3}$ }

\begin{document}

\maketitle

\begin{center}
{\it $^{1}$Joint Institute for Nuclear Research, 141980, Dubna, Moscow region, Russia}\\
{\it $^{2}$Skobeltsyn Institute of Nuclear Physics, Lomonosov Moscow State University, 119991, Moscow, Russia}\\
{\it $^{3}$ School of Physics and Astronomy, Sun Yat-sen University, Zhuhai 519082, China}

\end{center}

\vspace{0.5cm}

\begin{center}

{\bf Abstract }

\end{center}

\indent
To study the heavy quark production processes,
we use the transverse momentum dependent (TMD, or unintegrated)
gluon distribution function in a proton
obtained recently  using the Kimber-Martin-Ryskin
prescription from
the Bessel-inspired behavior of parton densities at small Bjorken $x$ values.
  We obtained a good agreement of our results with the latest HERA experimental data for reduced 
  cross sections $\sigma^{c\overline{c}}_{\rm red}(x,Q^2)$ and $\sigma^{b\overline{b}}_{\rm red}(x,Q^2)$, and also for deep inelastic
  structure functions  $F_2^c(x,Q^2)$ and $F_2^b(x,Q^2)$ in a wide range of $x$ and $Q^2$ values.
  Comparisons with the predictions based on the Ciafaloni-Catani-Fiorani-Marchesini evolution equation
  and with the results of conventional pQCD calculations performed at first three 
  orders of perturbative expansion are presented.
  


\vspace{1.0cm}


\newpage

\section{Introduction} \indent

Recently, important new data have appeared on the cross sections for the
open charm and beauty production in neutral current
deep inelastic
electron-proton scattering (DIS)
obtained\cite{H1+ZEUS:2018} by combining the results of
H1
and ZEUS
Collaborations at HERA.
Measurements
have shown that heavy flavour production in DIS proceeds predominantly via the photon-gluon fusion
process, $\gamma g \to Q\overline{Q}$, where $Q$ is the heavy quark. The cross section therefore depends strongly on
the gluon distribution in the proton and heavy quark mass. Moreover, an analysis
of the data in the framework of perturbative Quantum Chromodynamics (QCD) has been 
done\cite{H1+ZEUS:2018}, where the massive
fixed-flavour-number scheme
and different implementations of the variable-flavour-number scheme
were used.

The theoretical description of the heavy quark production processes can also be performed with transverse momentum dependent
(TMD), or non-integrated, functions of the density of partons (quarks and/or gluons) in a proton \cite{1,Abdulov:2021ivr}.
These quantities, depending on the fraction of the longitudinal momentum $x$ of the proton carried by the parton, the two-dimensional
transverse momentum of the parton ${\mathbf k_T^2}$, and the hard scale $\mu^2$ of a complex process, contain nonperturbative information
about the proton structure, including the transverse momentum.
The TMD factorization theorems provide the necessary basis for separating hard parton physics, described in terms of perturbative QCD,
and soft hadron physics.
Currently, there are a number of factorization approaches that include the dependence of the parton distribution fnctions (PDFs) on
the transverse momentum: for example, the Collins-Soper-Sterman \cite{2} approach,
developed for semi-inclusive processes with a finite and nonzero ratio between the rigid scale $\mu^2$ and total energy $s$, as well as
the approach to high-energy factorization \cite{3,3a} (or $k_T$-factorization \cite{4}), valid at a fixed limit of the hard scale
and at high energies.

With high-energy factorization, the TMD density of gluons satisfies the Balitsky-Fadin-Kuraev-Lipatov (BFKL) \cite{5} or
Ciafaloni-Catani-Fiorani-Marchesini (CCFM) \cite{6} evolution equations, which resum the contributions of large logarithms proportional to
$\alpha_s^n \ln^ns \sim \alpha_s^n \ln^n 1/x$.  These terms are important at high energies $s$ (or, equivalently, at low $x$ values).
Thus, high-order radiative corrections can be effectively taken into account in the estimated cross sections (namely, the part of the next-to-leading order
(NLO) + of the next-to-leading order (NNLO) + ... terms corresponding to the emission of the original gluons).
Phenomenological applications of the high-energy factorization approach augmented by the CCFM are well known in the literature (see, for example,
\cite{7,8,9,10,11,12,13,14,15,16,17} and references therein).

In addition to the CCFM equation, there are also other approaches to determining the TMD PDFs in a proton. They can be estimated using schemes based on the
usual Dokshitser-Gribov-Lipatov-Altarelli-Parisi (DGLAP) \cite{18} equations, namely the parton branching approach (PB) \cite{19} and
the Kimber-Martin-Ryskin (KMR) recipe \cite{21}.
The first of them gives a numerical iterative solution of the DGLAP evolution equations for collinear and TMD PDFs using the concept of
resolvable and non-resolvable branching and the application of Sudakov's formalism to describe the evolution of a parton from one scale to another
without decidable branching.
The splitting kinematics at each branching vertex is described by the DGLAP equations and the angular ordering condition for parton emission, which can be
used instead of the usual DGLAP ordering by virtuality.
The KMR approach is a formalism invented to construct TMD PDFs  from well-known traditional (collinear) PDFs under the key assumption that the dependence
of parton distributions on transverse momentum comes only in the last stage of evolution.
It is believed that the KMR procedure effectively takes into account a main part of the following
next-to-leading logarithmic (NLL) terms $\alpha_s (\alpha_s \ln \mu^2)^{n-1}$
compared to the leading logarithmic approximation (LLA), where terms are proportional to $\alpha_s^n \ln^n \mu^2$ taken into account only.
The KMR approach is currently being explored in NLO \cite{22} and is commonly used in phenomenological applications (see, for example,
\cite{8,10,11,12,13,14,15,16} and references therein), where the usual PDFs (for example, the NNPDF \cite{23} or CTEQ \cite{24} ones)
accepted numerically as input.
The relationship between PB and KMR scenarios has been discussed \cite{25}, and the relationship between PB and CCFM approaches has been recently
created \cite {26}.

The KMR formalism was used in our recent article \cite{KLSZ} for analytical calculations of the TMD PDFs.
These calculations are based on the \cite{Cvetic1, Q2evo, HT} expressions for usual PDFs obtained with the {\it generalized}
double asymptotic scaling (DAS) approach \cite{Munich, Q2evo, HT}.
This scaling, related to the asymptotic behavior of DGLAP evolution, was discovered many years ago by \cite{Rujula}. It was shown \cite{Cvetic1, Q2evo, HT} that
the flat initial conditions for the DGLAP equations used in the {\it generalized} DAS scheme lead to the Bessel-like behavior of PDFs at small $x$ values.
Using the results \cite{Cvetic1, Q2evo, HT}, analytical expressions for the TMD quark and gluon densities has be obtained \cite {KLSZ} in the leading order (LO)
\footnote{The obtained TMD PDFs are now implemented in the \textsc{tmdlib} package\cite{Abdulov:2021ivr}
  and are publicly available. Moreover, theyr are included in the Monte Carlo  event generator \textsc{pegasus}\cite{pegasus}.}.
In Ref.  \cite{KLSZ}, we have implemented various kinematic constraints that exist in the KMR recipe (namely, angular and strict ordering conditions),
and investigated the relationship between the differential and integral formulations of the KMR procedure recently mentioned in \cite{Golec-Biernat:2018hqo}.

In the present paper we continue our investigations and
analyze the combined H1 and ZEUS experimental data \cite{H1+ZEUS:2018} 
for the (reduced)
charm and beauty cross sections $\sigma^{c\overline{c}}_{\rm red}$ and $\sigma^{b\overline{b}}_{\rm red}$
and charm and beauty
contributions to the 
proton
structure functions (SFs) $F_2(x,Q^2)$ and $F_L(x,Q^2)$\cite{69,70,71} and their ratio 
for different $Q^2$ values.
Studying the earlier data on the charm SF $F^c_2$ in the proton from  H1\cite{Adloff:1996xq} and ZEUS\cite{Breitweg:1997mj}
  Collaborations at HERA, obtained for $x \sim 10^{- 4}$, it was found that the charm contribution to the total proton SF $F_2$ is about 25$\%$,
  which is significantly larger than that found by the European Muon Collaboration at CERN\cite{Aubert:1982tt} for large $x$, where SF $F^c_2$
  was only $1\%$ of $F_2$. Such a large value of $F^c_2$ required extensive experimental and theoretical studies of heavy quark production
  processes (see, for example, the 
  data
  \cite{H1+ZEUS:2018} studied in this paper, as well as the experimental data\cite{Aaij:2013mga} from LHCb Coolaboration at CERN for
  the prompt charm production in $pp$ collisions).
  Theoretical studies usually serve to confirm that HERA and LHC data can be described by perturbative charm generation within QCD (see, for example,
  reviews\cite{CooperSarkar:1997jk,Zenaiev:2016kfl,Gao:2017yyd} and references therein). 
We also note here that, historically, it was in the study of these processes that the $k_T$-factorization was introduced and
  tested (see\cite{3,3a,4}).

The production of charmed mesons ah hadron colliders is dominated by the $gg \to c\overline{c}$ subprocess, and therefore
  it provides a sensitive probe of the gluon density at small $x$. In particular the data\cite{Aaij:2013mga} of LHCb Collaboration provides
  an information on the gluon at values of $x$ as small as $x \sim 10^{-6}$ (see\cite{Cacciari:2015fta} and discussions therein).
  This very small-$x$ region is also crucial for the calculations
  of signal and backgroud processes for ultra-high energy neutrino astrophysics
  (see\cite{CooperSarkar:2011pa} and \cite{Gauld:2015kvh} for calculations of the high energy neutrino cross-section
  and the prompt atmospheric neutrino flux,
respectively).
A
  study of the heavy quark production will be contitued at future lepton-hadron and hadron-hadron
  colliders, such as LHeC, FCC-eh and FCC-hh, respectively (for a review, see\cite{AbelleiraFernandez:2012cc,Mangano:2016jyj}
 and references therein).

To study the process of the heavy quark production, we
produce the $k_T$-factorization predictions
in two ways, namely, using the framework of DAS approach 
and CCFM evolution equation.
The direct comparison of these predictions
is interesting and
could be rather useful to discriminate between the different 
approaches to evaluate the TMD parton (mainly gluon) density in a proton.
Then, we calculate the high-energy asymptotics of the 
heavy quark parts of the SFs $F_2$ and $F_L$   
at first three orders of perturbation theory
and present the numerical comparison of these higher-order 
predictions with corresponding results of the $k_T$-factorization calculations.

The outline of our paper is following. In Sections~2  and 3 we briefly describe
our theoretical input.
Section~4 presents our numerical results for the reduced
charm and beauty cross sections and charm and beauty parts of SFs
$F_2(x,Q^2)$ and $F_L(x,Q^2)$ 
in a wide $Q^2$ range.
Section~5 contains our conclusions.
In Appendix A we present the high energy asymptotics of the
  heavy quark contribution to the SFs $F_2$ and $F_L$
  at first three orders of perturbation theory. Appendix B contains the simple approximations of these formulas 
  for the ratio of the heavy quark parts of the SFs $F_2$ and $F_L$, which
  could be useful for subsequent applications.



\section{
$k_T$-dependent Wilson coefficient functions
} \indent

The  differential cross section $\sigma^{Q \bar Q}$ (hereafter $Q = c,b$) can be presented in
the simple form:
\begin{equation}
  {d^2 \sigma^{Q \bar Q} \over dx dy} = {2 \pi \alpha^2 \over x Q^4} \left[ \left( 1 - y + {y^2\over 2}\right) F_2^{Q}(x,Q^2) - {y^2 \over 2} F_L^{Q}(x, Q^2)\right],
\label{sigmaff}
\end{equation}
\noindent
where $F_k^{Q}(x, Q^2)$ (hereafter $k=2,L$)
are heavy quark parts of 
the proton SFs
$F_k(x, Q^2)$,
$x$ and $y$ are the usual Bjorken scaling variables.
Here we present the basic elements of the relations between SFs $F_2^{Q}(x,Q^2)$ and $F_L^{Q}(x,Q^2)$
and TMD PDFs. More detail analysis can be found in\cite{62}. 


In the $k_T$-factorization
approach, the SFs $F^Q_{k}(x,Q^2)$ are driven at small 
$x$ by gluons and related in the following way to the TMD gluon
distribution $f_g(x,{\mathbf k}^2_T,\mu^2)$: 
\begin{eqnarray}
F^Q_{k}(x,Q^2) = \int^1_{x} \frac{dx'}{x'} \int 
\frac{dk^2_{\bot}}{k^2_{\bot}} 
\; C_{k,g}(x,Q^2,m^2,k^2_{\bot},\mu^2)~ f_g\left(\frac{x}{x'}, k^2_{\bot},\mu^2\right).
 \label{d1}
\end{eqnarray}
\noindent
The functions $C_{k,g}(x,Q^2,m_f^2,k^2_{\bot},\mu^2)$ 
may be regarded as the structure 
functions of
the off-shell gluons with virtuality $k^2_{\bot}$ (hereafter we call them as
{\it Wilson coefficient functions}).
Following\cite{62}, we do not
use the Sudakov decomposition, which is sometimes quite convenient in high-energy 
calculations.
%
Here we only note that the property
$k^2=-k^2_{\bot}$
comes from the fact that
the Bjorken parton variable $x$ in the standard and in the
Sudakov approaches coincide. 

The $k_{\bot}$-dependent Wilson coefficient functions have the following form:
\be
C_{k,g}(x) = \frac{1}{\tilde \beta^4} \left[ \tilde \beta^2 
  \overline{C}_{k,g}(x) -12 bx^2 \frac{q^{\alpha}q^{\beta}}{Q^2}  \tilde{C}_{k,g}(x) \right],~~ 
\label{cgk}
\ee
where $\overline{C}_{k,g}(x)$ and $\tilde{C}_{k,g}(x)$ corresponds to application of the 
Feynman $P_F^{\alpha\beta}$ polarization tensor and additional tensor
of gluon polarization $P_a^{\alpha\beta}$:
\be
\hat P_F^{\alpha\beta}=-\frac{1}{2} \, g^{\alpha\beta},~~ \hat P_a^{\alpha\beta} =
\frac{6bx^2}{\tilde \beta^2} \,
\frac{q^{\alpha}q^{\beta}}{Q^2}.
\label{PolTe}
\ee
Hereafter
\be \tilde \beta^2=1-4bx^2,~~b=-\frac{k^2}{Q^2} \equiv  \frac{k_{\bot}^2}{Q^2},~~
a=\frac{m^2}{Q^2},~~ Q^2>0,  \, 
\label{tbeta}
\ee
\noindent
and we omitted the dependence of the coefficient functions on the heavy quark mass $m$, $Q^2$, $k^2_{\bot}$ and 
hard scale $\mu^2$.
The sum of the  polarizations tensors produces so-called BFKL polarizations tensor:
\be
\hat P^{\alpha\beta}_{BFKL}= 
\frac{k_{\bot}^{\alpha}k_{\bot}^{\beta}}{k_{\bot}^2}
=\frac{x^2}{-k^2}
p^{\alpha}p^{\beta}.~~~ 
\label{1dd}
\ee
Indeed, as it was shown in\cite{62}
\be
\hat P^{\alpha\beta}_{BFKL} ~=~ -\frac{1}{2}
\frac{1}{\tilde \beta^4} \left[\tilde \beta^2 g^{\alpha \beta} 
  -12 bx^2 \frac{q^{\alpha}q^{\beta}}{Q^2} \right] = \hat P_F^{\alpha\beta} + \hat P_a^{\alpha\beta}
\, .
\label{3dd} 
\ee
 
\noindent
After applying the photon polarizations tensors
 \begin{eqnarray}
\hat P^{(1)}_{\mu\nu} = -\frac{1}{2} g_{\mu\nu}~~\mbox{ and }~~
\hat P^{(2)}_{\mu\nu} = 4x^2\frac{k_{\mu}k_{\nu}}{Q^2} \, ,
\label{Photon}
 \end{eqnarray}
 which can be rewritten as
   \be
P^{(i)}_{\mu\nu} F_{\mu\nu} =
 \frac{e_f^2 a_s}{\beta^2}
 \;  f^{(i)} 
 , \, i = 1, 2 \, ,
 \label{fi}
\ee
the results for the coefficient functions $C_{k,2}(x)$
have been calculated in \cite{62}:
  \begin{eqnarray}
    C_{2,g}(x) &=& \frac{e_Q^2 a_s}{\beta^2} 
\;
\left[
f^{(1)} + 
\frac{3}{2\tilde \beta^2}\; f^{(2)} \right] \, \Theta(x_1-x)
 \label{3}\\
 C_{L,g} (x) &=& \frac{e_Q^2 a_s}{\beta^2}
\;
\left[
4bx^2 f^{(1)} + 
\frac{(1+2bx^2)}{\tilde \beta^2}\; f^{(2)} \right] \, \Theta(x_1-x)
\nonumber \\
&=&
 \frac{e_Q^2 a_s}{\beta^2}
\;
f^{(2)}\, \Theta(x_0-x) +4bx^2 \,
C^g_2 ,
 \label{4}
 \end{eqnarray}
  where $a_s=\alpha_s/(4\pi)$ is the strong coupling constant, $\Theta(x_1-x)$ is the Heaviside step function and 
 \be
x_1~=~\frac{1}{1+4a+b} \, .
 \label{1A}
 \ee
  
\noindent
Following to the representation (\ref{3dd}) of gluon polarization tensors, we can represent the
basic functions $f^{(i)}$ $(1,2)$ in the following form:
\be
f^{(i)}= 
\frac{1}{\tilde\beta^4}
\left[ \tilde \beta^2 \overline{f}^{(i)}  ~-~
  3bx^2 \tilde{f}^{(i)}\right] \, ,
\label{4dd}
\ee
where
\begin{eqnarray}
\overline{f}^{(1)} &=& -2 \overline{\beta} \Biggl[ 1 - \biggl(1-2x(1+b-2a) \; [1-x(1+b+2a)] 
\biggr) \; f_1  
\nonumber \\
    &+& (2a-b)(1-2a)x^2 \; f_2  \Biggr],
 \label{5}\\
\overline{f}^{(2)} &=& 8x\; \overline{\beta} \Biggl[(1-(1+b)x)  
-2x \biggl(bx(1-(1+b)x)(1+b-2a) + a\tilde \beta^2 \biggr)\; f_1  
\nonumber \\
    &+& bx^2(1-(1+b)x) (2a-b) \; f_2  \Biggr]
\label{6} 
\end{eqnarray}
and
\begin{eqnarray}
\tilde f^{(1)} &=& - \overline{\beta} \Biggl[ \frac{1-x(1+b)}{x}  
-2 \biggl(x(1-x(1+b))(1+b-2a) +a \tilde \beta^2 \biggr) \; f_1  
\nonumber \\
    &-& x(1-x(1+b))(1-2a) \; f_2  \Biggr],
 \label{5dd}\\
\tilde f^{(2)} &=& 4 \; \overline{\beta} ~(1-(1+b)x)^2 \Biggl[2   
- (1+2bx^2)\; f_1  
-bx^2 \; f_2  \Biggr].
\label{6dd} 
\end{eqnarray}
\noindent
Here
\be
\overline{\beta}^2=1-\frac{4ax}{(1-(1+b)x)}
\label{beta}
\ee
and
\be
f_1=\frac{1}{\tilde{\beta} \overline{\beta}} \;
\ln\frac{1+\overline{\beta} \tilde \beta}{1-\overline{\beta} \tilde \beta},
~~f_2=\frac{-4}{1-\overline{\beta}^2 \tilde \beta^2}.
\label{fi}
\ee


\subsection{
 The case of on-shell gluons 
} \indent

In the particular case of off-shell initial gluons, when $k^2 = 0$, we have (see\cite{62} for more details):
 \be
   C_{k,g}(x) ~=~ e_Q^2  a_s(\mu^2) \, B^{(0)}_{k,g}(x,a)
   \label{6a}
   \ee
   with
   \bea
    B^{(0)}_{2,g}(x,a) = x \,
\;
\left[
f^{(1)}(x,a) + 
\frac{3}{2}\; f^{(2)}(x,a) \right],~~
B^{(0)}_{L,g}(x,a) ~=~ x \, f^{(2)}(x,a),
\label{6.1}
 \end{eqnarray}
and
\begin{eqnarray}
f^{(1)}(x,a) &=& -2 \; \beta \Biggl[ \biggl(1-2x(1-x)(1-2a) \biggr) - 
\biggl(1-2x(1-2a)+2x^2(1-4a^2) \biggr) 
\;
L(\beta)
\Biggr],
\nonumber \\
& & \nonumber \\
f^{(2)}(x,a) &=& 8x \; \beta \Biggl[(1-x) 
  -2xa \;
  L(\beta)
\Biggr],  
\label{6.3} 
\end{eqnarray}
where
\be
\beta^2=\overline{\beta}^2(b=0)=1-\frac{4ax}{(1-x)},~~
L(\beta) = \frac{1}{\beta} \,
\ln\frac{1+\beta }{1-\beta } \, .
\label{Lbe}
\ee
Here $B^{(0)}_{k,g}(x,a)$ is the LO collinear Wilson coefficient function.
Equations (\ref{6.1}) --- (\ref{6.3}) 
coincide with the results\cite{Witten}.
Indeed, we have
 \begin{eqnarray}
B^{(0)}_{2,g}(x,a) &=&  -2x\beta \,
\Biggl[ \biggl(1-4x(2-a)(1-x) \biggr)
\nonumber \\
&-& 
\biggl(1-2x(1-2a)+2x^2(1-6a-4a^2) \biggr) \;
L(\beta) 
\Biggr]
\label{6.6}\\
B^{(0)}_{l,g}(x,a) &=& 8x^2\beta \,
\Biggl[ (1-x)
  - 2xa \;
  L(\beta)
\Biggr] \, ,
 \label{6.7}
 \end{eqnarray}
which are in agreement with the old results \cite{Witten}.

\section{
  Kimber-Martin-Ryskin approach} \indent

Here we present the main elements of TMD PDFs,
based on the KMR prescription in so-called interal formulation (see\cite{Golec-Biernat:2018hqo}) and 
DAS appoach for usual PDFs. More details, including the differential  formulation of the KMR prescription, can be found 
in our previous paper\cite{KLSZ}.

The TMD quark and gluon distributions (hereafter  $a = q, g$)
\be
f_a(x,k^2,\mu^2)
= T_a(\mu^2,k^2) \, \sum_{a'} \int\limits^{x_0}_x \, \frac{dz}{z} \,
P_{aa'}(z,k^2) \,  D_a\left(\frac{x}{z},k^2 \right) \, ,~~x_0=1-\Delta \, ,
\label{Def2}
\ee
where $D_a(x,\mu^2)$ are the conventional PDFs, $f_a(x,\mu^2) = x D_a(x,\mu^2)$,
$T_a(\mu^2,k^2)$ are the Sudakov form factors
and $P_{a a^\prime}(z,\mu^2)$ are the DGLAP splitting functions (see, for example,~(2.56) --- (2.60) in\cite{Buras:1979yt}):
\be
P_{aa'}(z,\mu^2) = 2a_s(\mu^2) \, P_{aa'}^{\rm (LO)} (z) + ...
\, .
\label{spliting}
\ee
\noindent

\subsection{Sudakov form factors $T_a(\mu^2, k^2)$} \noindent

The Sudakov form factor $T_a(\mu^2,k^2)$ have the following form (see~(2.4) in\cite{Golec-Biernat:2018hqo}):
\be
T_a(\mu^2,k^2)= \exp \left\{ - \int\limits^{\mu^2}_{k^2} \, \frac{dp^2}{p^2} \,   \sum_{a'} \int\limits^{x_0}_0 \, dz \, z P_{a'a}(z,k^2) \right\}\, .
\label{Ta}
\ee


\noindent
When $\Delta$ is a constant, we have
\be
T_a(\mu^2,k^2) = \exp \Bigl[ -d_a R_a(\Delta) s_1 \Bigr] \, ,
\label{Ta.1}
\ee
where
\bea
\z s_1=\ln \left( \frac{a_s(k^2)}{a_s(\mu^2)} \right),~ d_a=\frac{4C_a}{\beta_0},~ C_q=C_F,~ C_g=C_A,~
\beta_0=\frac{C_A}{3} \Bigl(11-2\varphi\Bigr),~\varphi=\frac{f}{C_A}=\frac{f}{3},
\nonumber \\
\z R_q(\Delta)= \ln\left(\frac{1}{\Delta}\right) - \frac{3x_0^2}{4},~
R_g(\Delta)= \ln\left(\frac{1}{\Delta}\right) - \left(1-\frac{\varphi}{4}\right) x_0^2
+ \frac{1-\varphi}{12} x_0^3 (4-3x_0)
\, .
\label{Ta.2}
\eea

\subsection{Conventional PDFs} \noindent

At LO, the conventional sea quark and gluon densities $f_a(x, \mu^2)$ 
can be written as follows 
\bea
f_a(x,\mu^2) &=& f^{+}_a(x,\mu^2) + f^{-}_a(x,\mu^2),  \nonumber \\
f^{+}_g(x,\mu^2) &=& \biggl(A_g + C \,
A_q \biggl)
		\overline{I}_0(\sigma) \; e^{-\overline d_{+} s} + O(\rho),~~~C=\frac{C_F}{C_A}=\frac{4}{9}, \nonumber \\
f^{+}_q(x,\mu^2) &=&
\frac{\varphi}{3} \,\biggl(A_g + C \,
A_q \biggl)
\tilde{I}_1(\sigma)  \; e^{-\overline d_{+} s}
+ O(\rho),
\nonumber \\
        f^{-}_g(x,\mu^2) &=& - C \,
A_q e^{- d_{-} s} \, + \, O(x),~~
	f^{-}_q(x,\mu^2) ~=~  A_q e^{-d_{-} s} \, + \, O(x),
	\label{8.02}
\end{eqnarray}

\noindent
where $C_A=N_c$, $C_F=(N_c^2-1)/(2N_c)$ for the color $SU(N_c)$ group,
$\overline{I}_{\nu}(\sigma)$ and
$\tilde{I}_{\nu}(\sigma)$ ($\nu=0,1$)
are the combinations of modified Bessel functions (at $s\geq 0$, i.e. $\mu^2 \geq Q^2_0$) and usual
Bessel functions (at $s< 0$, i.e. $\mu^2 < Q^2_0$):
\begin{equation}
\tilde{I}_{\nu}(\sigma) =
\left\{
\begin{array}{ll}
\rho^{\nu} I_{\nu}(\sigma) , & \mbox{ if } s \geq 0; \\
(-\tilde{\rho})^{\nu} J_{\nu}(\tilde{\sigma}) , & \mbox{ if } s < 0,
\end{array}
\right. \, ~~
\overline{I}_{\nu}(\sigma) =
\left\{
\begin{array}{ll}
\rho^{-\nu} I_{\nu}(\sigma) , & \mbox{ if } s \geq 0; \\
\tilde{\rho}^{-\nu} J_{\nu}(\tilde{\sigma}) , & \mbox{ if } s < 0,
\end{array}
\right.
\label{4}
\end{equation}
\noindent
where $\overline{I}_{0}(\sigma) = \tilde{I}_{0}(\sigma)$ and
\bea
&&s=\ln \left( \frac{a_s(Q^2_0)}{a_s(\mu^2)} \right),~~
a_s(\mu^2) \equiv \frac{\alpha_s(\mu^2)}{4\pi} = \frac{1}{\beta_0\ln(\mu^2/\Lambda^2_{\rm LO})},~~
\sigma = 2\sqrt{\left|\hat{d}_+\right| s
  \ln \left( \frac{1}{x} \right)},~~ \nonumber \\
&&\rho=\frac{\sigma}{2\ln(1/x)},~~
\tilde{\sigma} = 2\sqrt{-\left|\hat{d}_+\right| s
  \ln \left( \frac{1}{x} \right)},~~ \tilde{\rho}=\frac{\tilde{\sigma}}{2\ln(1/x)}
\label{intro:1a}
\eea
and
\begin{equation}
\hat{d}_+ = - \frac{4C_A}{\beta_0} = - \frac{12}{\beta_0},~~~
\overline d_{+} = 1 + \frac{4f(1-C)}{3\beta_0} =
1 + \frac{20f}{27\beta_0},~~~
d_{-} = \frac{4Cf}{3\beta_0}= \frac{16f}{27\beta_0}
\label{intro:1b}
\end{equation}
are the singular and regular parts of the anomalous dimensions, $\beta_0$
is the first coefficient of the QCD
$\beta$-function in the $\MSbar$-scheme.
The results for the parameters $A_a$ and $Q_0^2$ can be found in\cite{Cvetic1,Kotikov:2016oqm};
they were obtained
for $\alpha_s(M_Z)=0.1168$.

It is convenient to show the following expressions:
\begin{equation}
\beta_0 \, \hat{d}_+ = - 4C_A,~~~
\beta_0 \, \overline{d}_{+} = \frac{C_A}{3}\Bigl(11-2\varphi (1-2C)\Bigr),~~~
\beta_0 \, d_{-} = \frac{4Cf}{3}= \frac{4C_A \varphi}{3}.
\label{intro:1ba}
\end{equation}

\subsection{TMDs
  in KMR approach} \noindent

Now we can use~(\ref{Def2}) to find the results for TMDs without derivatives.
After some algebra we have
\bea
&&f_a(x,k^2,\mu^2)
= 4C_a \, a_s(k^2) T_a(\mu^2,k^2) \nonumber \\
&&\times \left(D_a(\Delta) \, f_a\left(\frac{x}{x_0},k^2\right) + D_a^+ f_a^+\left(\frac{x}{x_0},k^2\right) + D_a^{-} f_a^-\left({x\over x_0}, k^2\right) \right) \, .
\label{uPDF2.1}
\eea

\noindent
We can obtain (see more details in\cite{KLSZ}):
\bea
&&D_q(\Delta)= \ln\left(\frac{1}{\Delta}\right) - \frac{x_0}{4} (2+x_0),~
D_g(\Delta)= \ln\left(\frac{1}{\Delta}\right) - x_0 + \frac{x_0^2}{4} - \frac{x_0^3}{3},~
\nonumber \\
&&D_q^-(\Delta) = - \frac{x_0\varphi}{2} \,  \left(1-x_0 + \frac{2x_0^2}{3}\right),~
D_g^-(\Delta)=0,~
D_g^+= \frac{1}{\overline{\rho}_g} - x_0 + \frac{x_0^2}{4}
+ \frac{C\varphi}{3}, \,
\nonumber \\
&&D_q^+=  \frac{3x_0}{2C} \Biggl[\frac{1}{\overline{\rho}_s} \left(1 - x_0  + \frac{2x_0^2}{3}\right) -
\left( 1- \frac{x_0}{2} + \frac{2x_0^2}{9}  \right)\Biggr],
\label{uPDF2.3}
\eea
where
\be
\frac{1}{\rho_g} =
\frac{\overline{I}_1(\sigma)}{\overline{I}_0(\sigma)} = \frac{1}{\rho} \frac{I_1(\sigma)}{I_0(\sigma)},~~
 \frac{1}{\rho_q} =  \frac{1}{\rho} \, \frac{\overline{I}_0(\sigma)}{\tilde{I}_1(\sigma)}=
\frac{1}{\rho} \frac{I_0(\sigma)}{I_1(\sigma)}\, 
\label{rho.a}
\ee
and
\be
\overline{\sigma} =
\left\{
\begin{array}{ll}
\sigma\left(x \to \frac{x}{x_0}\right) , & \mbox{ if } s \geq 0; \\
\tilde{\sigma}\left(x \to \frac{x}{x_0}\right) , & \mbox{ if } s < 0.
\end{array}
\right. \, ,~~
\frac{1}{\overline{\rho}_a} = \frac{1}{\rho}_a \left(x \to \frac{x}{x_0}\right) \, .
\label{rho}
\ee

\subsection{Other prescriptions
} \noindent

{\bf 1.} For the phenomenological applications, we use
the {\it cut-off parameter} $\Delta$ in the angular ordering\cite{Golec-Biernat:2018hqo}
(the case of strong ordering can be found in \cite{KLSZ}):
\be
\Delta_{\rm ang}=\frac{k}{k+\mu} \, .
\label{Delta12}
\ee
\noindent
In all above cases, except the results for $T_a(\mu^2,k^2)$, we can simply replace the parameter $\Delta$
by 
$\Delta_{\rm ang}$.
For the Sudakov form factors, we note
that the parameters $\Delta$
contribute to the integrand
in (\ref{Ta}) and, thus, their
momentum dependence changes the results in (\ref{Ta.1}). To perform the correct evaluation of the integral (\ref{Ta}),
we should
recalculate the $p^2$ integration in (\ref{Ta}). So, we have
\be
T_a^{({\rm ang})}(\mu^2,k^2) = \exp \left[ -4C_a \int\limits^{\mu^2}_{k^2} \, \frac{dp^2}{p^2} \, a_s(p^2) R_a(\Delta_{\rm ang})  \right].
\label{Ta.2}
\ee

\noindent
The analytic evaluation of $T_a^{({\rm ang})}(\mu^2,k^2)$ is a very cumbersome procedure,
which will be
accomplished in the future. With the purpose of simplifying
our analysis, below we use the numerical
results for $T_a^{({\rm ang})}(\mu^2,k^2)$.\\

{\bf 2.}
As it was shown \cite{HT},
the results of the fits of the experimental data for SF $F_2$
are not very well at low $Q^2$ values.
To overcome this problem,
following to\cite{Cvetic1}, it is possible to use a modification of the
strong-coupling constant in the infrared region.
Specifically, usually it is possible to consider two modifications:
the ``frozen'' coupling constant (see, for example,\cite{Cvetic1,Badelek:1996ap}) and 
analytic one\cite{Shirkov:1997wi,Cvetic:2008bn}, which effectively increase the
argument of the strong coupling constant at small $\mu^2$ values, in accordance
with\cite{Kotikov:1993yw,DoShi}.
As one can see\cite{Cvetic1,Kotikov:2016oqm,Kotikov:2004uf},
the fits based on the "frozen" and
analytic strong coupling constants are rather
similar and describe the $F_2(x,Q^2)$ data in
the small $Q^2$ range significantly better than the canonical fit.

However, as it was shown in\cite{KLSZ},
the ``frozen'' coupling constant leads to a better agreements with data sets.
Thus, we will use it
in our present analysis. So,
we introduce a freezing
of the strong coupling constant by changing its argument as
$\mu^2 \to \mu^2 + M^2_{\rho}$, where $M_{\rho}$ is the $\rho$ meson mass\cite{Badelek:1996ap}.
Thus, in the formulae of Section~3
we introduce the following replacement
\begin{equation}
\alpha_s(\mu^2) \to \alpha_{\rm fr}(\mu^2)
= \alpha_s(\mu^2 + M^2_{\rho}) \, .
\label{Intro:2}
\end{equation}

\vskip 0.5cm

%


{\bf 3.}
In the phenomenological applications (see Section~4) the calculated TMD parton
densities will be used to predict the reduced cross sections $\sigma^{Q \bar Q}_{\rm red}$ 
and proton SFs $F_k^Q(x,Q^2)$.
%
According to $k_T$-factorization theorem\cite{3,4}, the theoretical 
predictions for these observables
can be obtained by convolution (\ref{d1})
of the TMD gluon densities
and corresponding off-shell production amplitudes. So, we need the TMD
quark and gluon distributions
in rather broad range of the $x$ variable, i.e. beyond the standart low $x$ range ($x \leq 0.05$).

It was shown (see\cite{KLSZ} and discussions therein) that
the analytic expressions for TMD parton densities can be modified in the form:
\be
f_a(x,k^2,\mu^2) \to f_a(x,k^2,\mu^2) \, \left(1-\frac{x}{x_0}\right)^{\beta_a(s)} \label{uPDF2.over1},
\ee
that in agreement with a similar modifications of conventional PDFs
(see, for example, the recent paper\cite{Kotikov:2018cju}, where similar extension has been done in the case of
EMC effect from the study of shadowing\cite{Kotikov:2017mhk} at low $x$ to antishadowing effect at $x \sim 0.1 - 0.2$).
The value of $\beta_a(0)$ can be estimated from the quark counting rules\cite{Matveev:1973ra}:
\be
\beta_v(0) \sim 3,~~  \beta_g(0) \sim \beta_v(0)+1 \sim 4,~~  \beta_q(0) \sim \beta_v(0)+2 \sim 5 \, ,
\label{bV}
 \ee
where the symbol $v$ marks the valence part of quark density.
Usually the $\beta_v(0)$, $\beta_g(0)$, $\beta_q(0)$ are determined from fits of experimental data
(see, for example,\cite{Jimenez-Delgado:2014twa,Krivokhizhin:2005pt}).
In our analysis, we use the numerical values of $\beta_{g}(0)=3.03$ which have been extracted\cite{KLSZ}
from the fit
to the inclusive $b$-jet production data taken by the CMS\cite{59} and ATLAS\cite{60} Collaborations
in $pp$ collisions at $\sqrt s = 7$~TeV.

\section{Phenomenological applications} \noindent

We are now in a position to apply the
TMD parton densities, obtained in\cite{KLSZ} and shown above, for phenomenological applications.
In the present paper we
consider the reduced charm and beauty cross sections $\sigma^{c\bar c}_{\rm red}$ and $\sigma^{b\bar b}_{\rm red}$
and charm and beauty contributions to the deep inelastic proton SFs
$F_2(x,Q^2)$, which are directly related with the
gluon content of the proton. These observables
were measured in $ep$ collisions at HERA with rather 
good accuracy (see\cite{H1+ZEUS:2018} and\cite{69,70,71}, respectively.) 
Additionally, in the evaluations below 
we will use latest TMD gluon density in a proton, obtained 
from the numerical solution of the CCFM evolution equation, namely, JH'2013 set 2 one\cite{JH}.
Our choice is motivated mainly by the fact that the CCFM equation
smoothly interpolates between the small-$x$ BFKL gluon dynamics and 
conventional DGLAP one, as it was mentioned above.
The input parameters of starting (initial) gluon distribution
implemented into the JH'2013 set 2 were fitted to describe the 
high-precision DIS data on structure functions $F_2(x, Q^2)$ 
and $F_2^c(x, Q^2)$ at $x \le 5 \cdot 10^{-3}$ (see\cite{JH} for more information).
Everywhere below, we set the charm and beauty masses 
to be equal to $m_c = 1.65$~GeV and $m_b = 4.78$~GeV\cite{63}.
We use the one-loop formula for the QCD coupling $\alpha_s$ 
with $n_f = 4$ quark flavours at $\Lambda_{\rm QCD} = 143$~MeV
(that corresponds to $\alpha_s(m_Z^2) = 0.1168$)
for for analytically calculated TMD gluon density as described above.
In the case of CCFM-evolved gluon,
we apply the two-loop expression for $\alpha_s$ with $n_f = 4$ 
and $\Lambda_{\rm QCD} = 200$~MeV, as it is dictated by the fit\cite{JH}.


\subsection{Reduced cross sections $\sigma^{Q\bar Q}_{\rm red}$ and SFs  
  $F_k^Q(x,Q^2)$
} \indent


Usually the differential cross section of heavy quark 
production in deep inelastic scattering (\ref{sigmaff})
are represented in terms of reduced cross sections $\sigma^{Q\bar Q}_{\rm red}$, 
which are defined as follows:
\be
    {d^2 \sigma^{Q \bar Q} \over dx dy} = {2 \pi \alpha^2 \over x Q^4} \, \left( 1 - y + {y^2\over 2}\right) \,
    \sigma^{Q \bar Q}_{\rm red} \, .
    \label{sigmaff.1}
\end{equation}
\noindent
Hence, taking together expressions~(\ref{sigmaff}) and~(\ref{sigmaff.1}),
the reduced cross section $\sigma^{Q \bar Q}_{\rm red}$ can be easily rewritten 
through $F_2^{Q}(x, Q^2)$ and $F_L^{Q}(x, Q^2)$ as
\be
 \sigma^{Q \bar Q}_{\rm red} =  F_2^{Q}(x,Q^2) - \frac{y^2}{1+(1-y)^2} \, F_L^{Q}(x, Q^2) = F_2^Q(x,Q^2) \, \left(1-\frac{y^2}{1+(1-y)^2}\overline{R}^Q(x,Q^2)\right) \, ,
\label{rsigmaff}
\end{equation}
\noindent
where the ratio $\overline{R}^Q(x,Q^2)$ in
defined as:
\begin{equation}
  \overline{R}^Q(x,Q^2) = \frac{F_L^Q(x,Q^2)}{F_2^Q(x,Q^2)} \, .
\label{eq:ori}
\end{equation}




%


\noindent
The evaluation below is based on the formulas~(\ref{rsigmaff}) and (\ref{d1})
with the coefficient functions as given by~(\ref{3}) --- (\ref{fi}).

Our numerical results for reduced cross sections $\sigma^{c\overline{c}}_{\rm red}$ and $\sigma^{b\overline{b}}_{\rm red}$
are shown in Figs.~1 and~2, respectively, in comparison with the latest H1 and ZEUS data \cite{H1+ZEUS:2018}.
The shaded bands represent the theoretical uncertainties of our calculations.
We find that the $k_T$-factorization predictions obtained 
using derived analytical expressions for TMD gluon density in a proton
are in perfect agreement with the
HERA data  
in a wide region of $x$ and $Q^2$ within the theoretical and experimental uncertainties,
both in normalization and shape.
These results tend to slightly overshoot the JH'2013 set 2 predictions in the 
region of small $x$ and especially at low $Q^2$.
At larger $Q^2$ and/or moderate or large $x \ge 10^{-2}$ 
the CCFM-evolved gluon density tends to overestimate the HERA data, 
that could be due to the determination of corresponding input parameters 
at small $x$ only (see\cite{JH}).
To estimate the scale uncertainties of our calculations,
the standard variations (by a factor of $2$) in default renormalization and factorization scales,
which were set to be equal to $\mu_R^2 = 4m_{Q}^2 + Q^2$
and $\mu_F^2 = Q^2$, respectively, were introduced.
To show the contribution of the longitudinal structure functions $F_L^c(x,Q^2)$ and $F_L^b(x,Q^2)$, 
we present also in Figs.~1 and~2 the results for $F_2^c(x,Q^2)$ and $F_2^b(x,Q^2)$ as dotted curves. 
The difference between the estimated 
$\sigma^{Q \bar Q}_{\rm red}$ and  $F_2^Q(x,Q^2)$ is coming from the 
contribution of the longitudinal SFs 
$F_L^Q(x,Q^2)$, as it can been clearly seen from~(\ref{rsigmaff}).
So, our calculations show that these contributions 
are rather important at low $x$.

To show the difference between $\sigma^{Q \bar Q}_{\rm red}$ and $F_2^Q(x,Q^2)$ more clearly, 
in Figs.~3 and~4 we present our 
results for SFs $F_2^c(x,Q^2)$
and $F_2^b(x,Q^2)$ in comparison with the latest ZEUS\cite{69} and
H1\cite{70,71} data. Our corresponding predictions for the reduced cross 
sections $\sigma^{c\overline{c}}_{\rm red}$ and
$\sigma^{b\overline{b}}_{\rm red}$ are presented here as dotted curves.
One can see again
that the results obtained with analytically evaluated TMD gluon density
are in good
agreement with the latest HERA data
for both structure functions $F_2^c(x,Q^2)$
and $F_2^b(x,Q^2)$ in a wide region of $x$ and $Q^2$.
The CCFM-evolved gluon JH'2013 set 2 provides a bit 
worse description of the HERA data, although these results are rather close to the measurements.
We find that the discrepancy between 
two considered approaches tends to be more clearly pronounced 
at large $Q^2$.




\subsection{
  Ratio $\overline{R}^Q(x,Q^2)$} \indent

Following the results of\cite{62} and using our coefficient functions obtained 
in Section~2 and TMD gluon density presented 
in Section~3, now we can produce predictions for the
ratio $\overline{R}^Q(x,Q^2)$ according to~(\ref{eq:ori}). 
Results of our calculations for $\overline{R}^c(x,Q^2)$ are presented in Fig.~5,
where we plot this ratio as a function of $x$ in a wide $Q^2$ range.
As earlier, we have applied two TMD gluon densities in a proton discussed above\footnote{The 
predictions for the ratio $\overline{R}^b(x,Q^2)$ are rather similar and not shown here.}.

Our calculations leads to more or less flat (independent on $x$) behaviour 
of $\overline{R}^c(x,Q^2)$ with $\overline{R}^c \sim 0.1$ at low $Q^2 \sim 5$~GeV$^2$ and
$\overline{R}^c \sim 0.3 - 0.4$ at higher $Q^2 \sim 200$~GeV$^2$ values.
The results 
obtained with our TMD gluon and CCFM-evolved one are in a good agreement to each other.
The difference between them is visible at very large $Q^2$ only.
Moreover,
the
  obtained predictions are in good agreement 
  with ones\cite{62}, which were based in-turn on rather old representations 
  for TMD gluon density (see\cite{OldBlu} and more recent\cite{Abdulov:2021ivr}). 

  As a next point of our study,
  we would like to compare the results for 
  the ratio  $\overline{R}^c(x,Q^2)$,
  obtained in $k_T$-factorization 
  with the one $\hat{R}^c(x,Q^2)$ (see \ref{eq:rio}),
  where the ratio $\hat{R}^c(x,Q^2)$ 
  was obtained in the conventional (collinear) QCD factorization
  at first three orders of perturbation theory (see Appendix A) represented by 
  the solid, dashed and dotted grey curves in Fig.~5.  
  To evaluate the latter, we have used the LO DAS parton densities presented in Section 3.2.
  Note that the results for this ratio, obtained using the power-like behaviour $x^{-\Delta}$ of the collinear PDFs,
  can be found in\cite{Ivanov:2008era}.
Our calculations show that 
the $k_T$-factorization predictions rather close to the results obtained beyond LO of collinear
perturbation theory.
  This is in complete agreement with the usual statement about the property of $k_T$-factorization, which 
  resums the main part of
  higher order pQCD contributions at small $x$. Indeed, 
  the LO results obtained in the collinear perturbation theory lead to
too small values\footnote{An absence of $x$-dependence of the collinear LO results for
the ratio $\hat{R}^c(x,Q^2)$ is discussed in Appendices A and B.} for
the ratio $\hat{R}^c(x,Q^2)$.
However, the $Q^2$-dependence of the ratio $\hat{R}^c(x,Q^2)$ in the NLO and NNLO results are noticebly different 
from the
corresponding $Q^2$-dependence evaluated with the TMD gluons.
In fact, in the $k_T$-factorization approach the ratio $\overline {R}^c(x,Q^2)$ 
grows fastly when $Q^2$ increased whereas in collinear
perturbation theory the ratio $\hat{R}^c(x,Q^2)$ grows slowly.

Of course, the observed difference 
between the predicted $\overline {R}^c(x,Q^2)$ and $\hat{R}^c(x,Q^2)$ ratios
at moderate and large $Q^2$
is unclear and needs an explanation, especially because
there are no experimental data for the SF $F_L^c(x,Q^2)$ and, accordingly, for the ratio $\overline{R}^c(x,Q^2)$.
Indeed, the $k_T$-factorization with 
  the estimated $\overline{R}^c(x,Q^2)$  
  leads to good agreement between experimental data and theoretical predictions for both
  reduced cross sections $\sigma^{c\overline{c}}_{\rm red}$ and SF $F_2^c(x,Q^2)$, as one can see 
  in Figs.~1 and~3. From another side, 
  the experimental data for both $\sigma^{c\overline{c}}_{\rm red}$ and $F_2^c(x,Q^2)$
  are in good agreement with the corresponding theoretical predictions obtained 
  in the framework of collinear approach\cite{H1+ZEUS:2018,69,70,71} (see also section 2.5 in the recent review\cite{Gao:2017yyd}).
  However, we would like to note that there is a quite similar situation between the exclusive reduced 
  cross section $\sigma_{\rm red}(x,Q^2)$ and 
  $F_2(x,Q^2)$: experimental data for both of these observables 
  are in good agreement with the corresponding theoretical predictions (see\cite{Abramowicz:2015mha} and discussions therein), 
  where the calculated SF $F_L(x,Q^2)$ is known to be very sensitive to low $x$ resummation 
  (see, for example, Section 9.3 in the recent review\cite{Gao:2017yyd}). But, apart of $F_L^c(x,Q^2)$, 
  the SF $F_L(x,Q^2)$ is measured at the HERA (see\cite{Andreev:2013vha} and references therein). 
  Therefore, it seems that in order to understand the observed difference between
  the predictions for $\overline {R}^c(x,Q^2)$ and $\hat{R}^c(x,Q^2)$ ratios at large $Q^2$
  one have to investigate first the SF $F_L(x,Q^2)$ using the same approaches, that is 
  out of our present consideration.  
  We plan to perform such investigation in forthcoming study
  and then, having experience in the latter, return to the study 
  its heavy quark parts and, correspondingly, the ratio $\overline{R}^c(x,Q^2)$.

\section{Conclusions} \indent

We have studied the heavy quark production processes with using
the transverse momentum dependent
gluon distribution function in a proton
obtained recently\cite{KLSZ} using the Kimber-Martin-Ryskin
prescription from
the Bessel-inspired behavior of parton densities at small Bjorken $x$ values.
The
Bessel-like behavior of parton densities at small Bjorken $x$ was obtained\cite{Cvetic1,Q2evo,HT,Munich}
in-turn
in the case of the flat initial conditions for DGLAP evolution equations
in the double scaling QCD approximation.
To construct the TMD parton distributions we
implemented\cite{KLSZ} the different
treatments of kinematical constraint, reflecting the angular and strong ordering conditions
and discussed the relations between the differential and integral formulation of the KMR approach.
Additionally, we have tested the TMD gluon density in a proton obtained
from the numerical solution of the CCFM evolution equation, which smoothly 
interpolates between the small-$x$ BFKL dynamics and large-$x$ DGLAP ones.

We have considered  the (reduced) cross sections $\sigma^{Q \bar Q}_{\rm red}$ (where $Q = c,b$)
and charm and beauty contributions to the deep inelastic proton SFs
$F_2(x,Q^2)$ and $F_L(x,Q^2)$.
To show an importanse of the longitudinal structure function $F_L^c(x,Q^2)$ and $F_L^b(x,Q^2)$, we compare the results
for $\sigma^{c\overline{c}}_{\rm red}$ and  $\sigma^{b\overline{b}}_{\rm red}$ with the SFs and $F_2^c(x,Q^2)$ and $F_2^b(x,Q^2)$.
We achieved a good agreement between the HERA experimental data 
for these observables and our theoretical predictions 
and demostrated
the importance of the contributions of $F_L^c(x,Q^2)$ and $F_L^b(x,Q^2)$ at small $x$. 
Concerning the ratio of the proton SFs, namely, $\overline{R}^c(x,Q^2) = F_L^c(x,Q^2)/F_2^c(x,Q^2)$,
we show that the results of $k_T$-factorization calculations 
are similar to the ones obtained beyond LO of collinear
perturbation theory. This effect is clearly visible for $Q^2 \leq 12$ GeV$^2$. 
However, starting with $Q^2 \geq 12$ GeV$^2$, 
the $k_T$-factorization leads to larger values for the ratio $\hat{R}^c(x,Q^2)$, that needs an 
additional investigations.


As we discussed already in Section 4.2, as
the next step, we plan to study the longitudinal
structure function $F_L(x,Q^2)$ and compare our future results 
with the previous ones \cite{KLZ,Kotikov:2004uf} 
and \cite{Kotikov:1993yw,Kotikov:1994jh,Kaptari:2019lfj}, obtained in the framework of $k_T$-factorization and
collinear perturbation theory, respectively. 
This study is important in itself and will provide some kind of clue for solving the problem of
  differences in the predictions for the ratio $\overline{R}^c(x,Q^2)$ obtained in the
  framework of $k_T$-factorization approach and conventional (collinear) QCD factorization (see Section 4.2).


Moreover, we plan to extend the present analysis beyond the LO approximation.
We will obtain the results for the NLO TMD parton densities using the corresponding NLO results \cite{Cvetic1,Q2evo,HT}
for the standard PDFs in the generalized DAS approach. We will accept also the results for the NLO matrix elements
(see \cite{72,25} and references and discussions therein).
Such results are seems to be extremely important for future experiments,
in particular, for experiments at the Electron-Ion Collider (EIC) and Electron-Ion Collider in China (EiCC)
(see\cite{Accardi:2012qut,Anderle:2021wcy} and discussions and references therein).
So, at the EIC, an essential low $x$ (up to $x \sim 10^{-4}$) region is expected to be 
probed, thus providing us with new and precise data for DIS SFs, especially data for 
longitudinal SF $F_L(x,Q^2)$. The EiCC could provide a new information on the 
light and sea quark density in a proton, that is, of course, important to 
produce and update the theoretical high-order predictions for $F_L(x,Q^2)$.
Moreover, EIC and EiCC measurements could be important to distinguish between 
the different non-collinear QCD evolution scenarios widely discussed at 
present (see, for example, review \cite{1}).


\section*{Acknowledgements}
A.V.L. thanks H.~Jung, S.P.~Baranov and M.A.~Malyshev for very useful discussions
and remarks.
A.V.L. is grateful to
DESY Directorate for the support in the framework of
Cooperation Agreement between MSU and DESY on phenomenology of the LHC processes
and TMD parton densities.
 P.Z. is supported in part by the National Natural Science Foundation of China
(Grants No. 11975320).

\section*{Appendix A.  Collinear approach} \indent
\label{App:A}
\def\theequation{A\arabic{equation}}
\setcounter{equation}{0}


It is easily to obtain the following
results in the collinear generalized DAS approach (see \cite{Illarionov:2008be,IllaKo}):
\begin{equation}
  \hat{F}_k^Q(x,Q^2) = C_{k,g}(x,Q^2,m^2_Q) \, \otimes f_g(x,Q^2),
\label{eq:pm3}
\end{equation}
where the SFs, obtained in the generalized DAS approach, were marked as $\hat{F}_k^Q(x,Q^2)$. Here
$\otimes $ is the Mellin convolution:
\bea
C_{k,g}(x,Q^2,m^2_f) \, \otimes f_g(x,Q^2) &=& \int_{x/x_2}^1 \, \frac{dy}{y} \, C_{k,g}(x/y,Q^2,m^2_f) \, f_g(y,Q^2) \nonumber \\
&=&
\int_{x}^{x_2} \, \frac{dy}{y} \, C_{k,g}(y,Q^2,m^2_f) \, f_g(x/y,Q^2),
\label{otimes}
\eea
\noindent
where
 \be
x_2=x_1(b=0)~=~\frac{1}{1+4a_Q}. \,
 \label{x1}
 \ee


\noindent
It can be represented as
\begin{equation}
  \hat{R}^Q(x,Q^2) = \frac{\hat{F}_L^Q(x,Q^2)}{\hat{F}_2^Q(x,Q^2)} =
\frac{C_{L,g}(x,Q^2,m^2_Q) \, \otimes f_g(x,Q^2)}{C_{2,g}(x,Q^2,m^2_Q) \, \otimes f_g(x,Q^2)} \, ,
  \label{eq:rio}
\end{equation}
where $f_g(x,Q^2)$ is given in (\ref{8.02}).
In fact, the ratio $\hat{R}^Q(x,Q^2)$ depends slowly on
 non-perturbative input $f_g(x,Q^2)$, which contribute to the both numerator and denominator 
 of the ratio $\hat{R}^Q(x,Q^2)$. 

  Using the results of $k_T$-factorization and BFKL approach\cite{3,Catani:1996sc}
  (see also\cite{Catani:1992zc,Kawamura:2012cr}), the results for the high energy limit of collinear coefficient functions of the
  heavy quark production process in all orders of perturbation theory were obtained. Thus, using the 
  results\cite{Catani:1996sc}, below we give formulas for the high energy asymptotics of collinear coefficient functions of the
  heavy quark production process in the first three orders of the perturbation theory.
  
\subsection*{LO} \indent

Taking the LO Wilson coefficient~(\ref{6a}), 
results (\ref{6.6}) and (\ref{6.7}) for on-shell coefficient functions and
the PDFs considered
in the section 3.2,
we have for the  ratio $\hat{R}^Q_{\rm LO}(x,Q^2)$:
\begin{equation}
\hat{R}^Q_{\rm LO}(x,Q^2) =
\frac{B^{(0)}_{L,g}(x,Q^2,m^2_Q) \, \otimes f_g(x,Q^2)}{B^{(0)}_{2,g}(x,Q^2,m^2_Q) \, \otimes f_g(x,Q^2)} \, ,
  \label{eq:ri}
\end{equation}
where the dependence of gluon density $f_g(x,Q^2)$ should be rather week.
In fact, there is no $x$-dependence at all (see Fig.~5), which is 
associated with the property of the Mellin convolution
(\ref{otimes})  in the low $x$ region (see Appendix B). 

\subsection*{NLO} \indent

Through NLO, we have
\be
   C_{k,g}(x) ~=~ e_Q^2  a_s(\mu^2) \, \bigl[B^{(0)}_{k,g}(x,a) + a_s(\mu^2) \, B^{(1)}_{k,g}(x,a)\bigr] \, .
   \label{6aNLO}
   \ee
\noindent
The NLO coefficient functions $B^{(1)}_{k,g}(x,a)$ of photon-gluon fusion subprocess are rather lengthy and not published in
print; they are only available as computer codes\cite{Laenen:1992zk}.
Following \cite{Illarionov:2008be,IllaKo}, 
it is sufficient to work in the high energy
regime, defined by $x\ll1$, where they assume the compact form\footnote{Following Ref. \cite{Catani:1996sc}, we will use the case $M^2=4m^2$ in the colliner approach. We would like to note that in the
  original papers\cite{Catani:1992zc,Kawamura:2012cr} the scale $M^2=m^2$ has been used, which is unconsistent with the results in eqs.
  (\ref{eq:ja}), (\ref{eq:nlo1}) and (\ref{eq:nnlo1}).}\cite{Catani:1992zc,Catani:1992zc,Kawamura:2012cr}:
  \begin{equation}
  B_{k,g}^{(1)}(x,a)=\beta \bigl[R_{k,g}^{(1)}(1,a) + 4 C_A B_{k,g}^{(0)}(1,a) L_{\mu} \bigr] \, ,~~ L_{\mu}=\ln\frac{M^2}{\mu^2} \, , M^2=4m^2, \,
\label{eq:nlo}
\end{equation}
with
\bea
&&R_{2,g}^{(1)}(1,a)~=~\frac{8}{9}C_A[5+(13-10a)J(a)+6(1-a)I(a)],~~
\nonumber \\
&&R_{L,g}^{(1)}(1,a)~=~-\frac{16}{9}C_A x_2\bigl\{1-12a-[3+4a(1-6a)]J(a)+12a(1+3a)I(a)\bigr\}
\label{eq:nloA}
\eea
and
  \bea
   B_{2,g}^{(0)}(1,a) &=&  \frac{2}{3} \, [1+2(1-a)J(a)],
\nonumber\\
B_{L,g}^{(0)}(1,a) &=&  \frac{4}{3} \, x_2 \, [1+6a-4a(1+3a)J(a)],
\label{Bk0}
\end{eqnarray}
where\footnote{The functions $J(a)$ and $I(a)$ in (\ref{eq:ja}) and (\ref{eq:nlo1}) coincide with ones in 
  \cite{Illarionov:2008be} and differ of ones in \cite{Catani:1992zc,Kawamura:2012cr,Catani:1996sc} by an additional factor
  $4a$. The function $K(a)$ in (\ref{eq:nnlo1}) coincides with the combination $4a[K(a)+\ln(4ax_2)I(a)]$.}
\bea
&&J(a) = - \sqrt{x_2}\ln t,\qquad t=\frac{1-\sqrt{x_2}}{1+\sqrt{x_2}},
\label{eq:ja} \\
&&I(a)=-\sqrt{x_2}\left[\zeta(2)+\frac{1}{2}\ln^2t-\ln(ax_2)\ln t+2\Li_2(-t)\right]
\label{eq:nlo1}
\eea
with
\be
\Li_2(x)=-\int_0^1\,\frac{dy}{ y}\, \ln(1-xy)
\label{Li2}
\end{equation}
being the dilogarithmic function.
We would like to note that $B_{k,g}^{(0)}(1,a)$ are the first moments 
of the LO Wilson coefficients $B_{k,g}^{(0)}(x,a)$ (see~(\ref{6.6}) and~(\ref{6.7})):
\begin{equation}
B_{k,g}^{(0)}(n,a)=\int_0^{x_2}dx\,x^{n-2} B_{k,g}^{(0)}(x,a) \, .
\label{eq:melB}
\end{equation}
\noindent
So, at the NLO we have for the ratio $\hat{R}^Q(x,Q^2)$:

\begin{equation}
\hat{R}^Q_{\rm NLO}(x,Q^2) =
\frac{\bigl[B^{(0)}_{L,g}(x,a) + a_s(\mu^2) \, B^{(1)}_{L,g}(x,a)\bigr] \, \otimes f_g(x,Q^2)}{\bigl[B^{(0)}_{2,g}(x,a) + a_s(\mu^2) \, B^{(1)}_{2,g}(x,a)\bigr] \, \otimes f_g(x,Q^2)} \, ,
  \label{eq:tri}
\end{equation}
where the LO gluon density $f_g(x,Q^2)$ given by~(\ref{8.02}) is used because its contribution to the ratio $\hat{R}^Q_{\rm NLO}(x,Q^2)$ is
strongly suppressed (see Appendix B).

  
\subsection*{NNLO} \indent


Following to the results \cite{Kawamura:2012cr}, we have for the
coefficient function:
\be
   C_{k,g}(x) ~=~ e_Q^2  a_s(\mu^2) \, \bigl[B^{(0)}_{k,g}(x,a) + a_s(\mu^2) \, B^{(1)}_{k,g}(x,a) + a^2_s(\mu^2) \, B^{(2)}_{k,g}(x,a)\bigr] \, ,
   \label{6aNNLO}
   \ee
   where in the high energy regime the coeffcicient $B^{(2)}_{k,g}(x,a)$ has 
   the compact form:
\begin{equation}
  B_{k,g}^{(2)}(x,a)=\beta \, \ln(1/x) \,\bigl[R_{k,g}^{(2)}(1,a) + 4 C_A R_{k,g}^{(1)}(1,a) L_{\mu} + 8 C_A^2 B_{k,g}^{(0)}(1,a) L^2_{\mu} \bigr] + O(x^0)\, ,
\label{eq:nnlo}
\end{equation}
with
\bea
&&R_{2,g}^{(2)}(1,a)~=~\frac{32}{27}C_A^2 \, [46+(71-92a)J(a)+3(13-10a)I(a)-9(1-a)K(a)],~~\nonumber \\
&&R_{L,g}^{(1)}(1,a)~=~\frac{64}{27}C_A^2 x_2\bigl\{34+240a-[3+136a+480a^2]J(a)
+3[3+4a(1-6a)]I(a) \nonumber \\
&& \hspace{2cm} +18a(1+3a)K(a)\bigr\} \, ,
\label{eq:nnloA} 
\eea
where
$J(a)$ and $I(a)$ are defined by~(\ref{eq:ja}) and~(\ref{eq:nlo1}), respectively, and
\bea
  K(a)&=&-\sqrt{x_2}\biggl[ 4\Bigl(\zeta(3)+\Li_3(-t)-\Li_2(-t)\ln t-2S_{1,2}(-t) \Bigr)+ 2\ln(ax_2)
    \nonumber \\
    &&\times \Bigl(\zeta(2)+2\Li_2(-t)\Bigr) - \frac{1}{3}\ln^3t-\ln^2(ax_2)\ln t+\ln(ax_2)\ln^2 t\biggr],
\label{eq:nnlo1}
\eea
where $t$ is given in (\ref{eq:ja}) and
\be
\Li_3(x)=\int_0^1\,\frac{dy}{ y}\, \ln(y) \, \ln(1-xy),~~S_{1,2}(x)= \frac{1}{2} \, \int_0^1\,\frac{dy}{ y}\, \ln^2(1-xy),~
\label{Li3}
\end{equation}
 are the trilogarithmic function $\Li_3(x)$ and Nilsen Polylogarithm $S_{1,2}(x)$ (see\cite{Devoto:1983tc}).
The results for $K(a)$ in the form of harmonic Polylogarithms\cite{Remiddi:1999ew} can be found
in\cite{Kawamura:2012cr}.
 
So, at the NNLO we have for the ratio $\hat{R}^Q(x,Q^2)$:
\begin{equation}
\hat{R}^Q_{\rm NNLO}(x,Q^2) =
\frac{\bigl[B^{(0)}_{L,g}(x,a) + a_s(\mu^2) \, B^{(1)}_{L,g}(x,a) + a^2_s(\mu^2) \, B^{(2)}_{L,g}(x,a)\bigr] \, \otimes f_g(x,Q^2)}{\bigl[B^{(0)}_{2,g}(x,a)
    + a_s(\mu^2) \, B^{(1)}_{2,g}(x,a)+ a^2_s(\mu^2) \, B^{(2)}_{2,g}(x,a)\bigr] \, \otimes f_g(x,Q^2)} \, ,
  \label{eq:tri}
\end{equation}
where the LO gluon density $f_g(x,Q^2)$ given by~(\ref{8.02}) is used because
its contribution to the ratio $\hat{R}^Q_{\rm NNLO}(x,Q^2)$ is strongly suppressed
(see Appendix B).


 
\section*{Appendix B. Collinear results in DAS approach} \indent
\label{App:B}
\def\theequation{B\arabic{equation}}
\setcounter{equation}{0}

The use of the DAS approach for the PDFs makes it possible to significantly simplify the
  formulas for the relation $\hat{R}^Q(x,Q^2)$. We will show this below.

Taking the results (\ref{6.6}) and (\ref{6.7}) for on-shell coefficient functions and
the PDFs considered
in the section 3.2,
it is easily to obtain the following LO results in the generalized DAS approach (see\cite{Illarionov:2008be}):
\begin{equation}
  \hat{F}^Q_k(x,Q^2) =
  M_{k,g}(1,Q^2,m^2_Q) \, f_g(x,Q^2),
\label{eq:pm3}
\end{equation}
where
$M_{k,g}(1,Q^2,m^2)$ is the first Mellin moment ($n=1$) (see (\ref{eq:ja})),
where the Mellin moments can be defined as
\begin{equation}
M_{k,g}(n,Q^2,m^2_Q)=\int_0^{x_2}dx\,x^{n-2}C_{k}^g(x,Q^2,m^2_Q) \, , 
\label{eq:mel}
\end{equation}
where $x_2$ are given by~(\ref{x1}).
In fact, the non-perturbative input $f_g(x,Q^2)$ does cancel in the $\hat{R}_f$ ratio and we have
\begin{equation}
  \hat{R}^Q(x,Q^2) = \frac{\hat{F}_L^Q(x,Q^2)}{\hat{F}_2^Q(x,Q^2)} =
  \frac{M_{L,g}(1,Q^2,m_Q^2)}{M_{2,g}(1,Q^2,m_Q^2)},
\label{eq:riB}
\end{equation}
in the case, where the moments $M_{k,g}(1,Q^2,m_Q^2)$ have no singularities at $n \to 1$.

\subsection*{LO} \indent

Taking the integral (\ref{eq:mel}), which becomes to be (\ref{eq:melB})  at the LO, we can
obtain the
results (\ref{Bk0}), using
(see\cite{Illarionov:2008be})
 the following auxiliary formulas\footnote{In the original paper\cite{Illarionov:2008be} the second result in (\ref{eq:bet}) was presented with the error ``... $1-2a$ ...'' instead of the
  correct expression ``... $1+6a$ ...'' and the third result in (\ref{eq:Lbet}) was presented with the error ``... $x_2^2/3$ ...''
  instead of the  correct expression ``... $x_2^2/6$ ...''.}
 \begin{eqnarray}
\int_0^{x_2} \, dx \, x^m \beta&=& 
\left\{\begin{array}{ll}
1-2aJ(a), & \mbox{if}~m=0 \\
\frac{x_2}{2}[1+6a-4a(1+3a)J(a)], & \mbox{if}~m=1 \\
\frac{x_2^2}{3}[(1+3a)(1+10a)-6a(1+6a+10a^2)J(a)], & \mbox{if}~m=2 \\
\end{array}\right.,\qquad
\label{eq:bet}\\
\int_0^b \, dx \, x^m \, L(\beta) &=& 
\left\{\begin{array}{ll}
J(a), & \mbox{if}~m=0 \\
-\frac{x_2}{2}[1-(1+2a)J(a)], & \mbox{if}~m=1 \\
-\frac{x_2^2}{6}[3(1+2a)-2(1+4a+6a^2)J(a)], & \mbox{if}~m=2
\end{array}\right. \,
\label{eq:Lbet}
\end{eqnarray}

\noindent
So, at LO the small-$x$ approximation formula (\ref{eq:ri}) thus reads
\begin{equation}
  \hat{R}^Q_{\rm LO}(x,Q^2)=
2x_2\frac{1+6a_Q-4a_Q(1+3a_Q)J(a_Q)}{1+2(1-a_Q)J(a_Q)} \, ,
\label{eq:lo}
\end{equation}
which is $x$-independent, that is in full agreement with the numerical evaluation of the $\overline{R}^Q_{\rm LO}(x,Q^2)$ in (\ref{eq:ri}).

\subsection*{NLO} \indent

At
NLO, the coefficient function $C^g_{k}(x)$ has the form (\ref{6aNLO}) with the NLO coefficients $B^{(1)}_{k,g}(x,a)$ 
given by~(\ref{eq:nlo}) and (\ref{eq:nloA}). Its moments
$M_{k,g}(n,Q^2,\mu^2)$ exhibit the corresponding structure
\be
M_{k,g}(n,Q^2,\mu^2) = e_Q^2a_s(\mu^2) \, \bigl[B^{(0)}_{k,g}(n,a) + a_s(\mu^2) \, B^{(1)}_{k,g}(n,a)\bigr] \, .
\label{eq:exp}
\ee

\noindent
The Mellin transforms of $B_{k,g}^{(1)}(x,a)$
exhibit singularities in the limit $n \to 1$,
which lead to modifications in (\ref{eq:pm3}).
As was shown\cite{39},
the terms involving $1/\delta$ at  $n=1+\delta \to 1$ (which
correspond to singularities of 
the Mellin moments $M_{k,g}(n)$ (see (\ref{eq:mel}))
at $n \to 1$)
depend on the exact form of the asymptotic
low-$x$ behavior encoded in $f_g(x,\mu^2)$.
Using the results for $f_g(x,\mu^2)$ from (\ref{8.02}), we obtain the
modification is
simple modification (see\cite{Illarionov:2008be} and discussions therein)
\begin{equation}
  \frac{1}{\delta} \to \frac{1}{\tilde \delta_{\pm}},~~
\frac{1}{\tilde \delta_+(x)}
\approx \frac{1}{\rho_g(x,\mu)}\,,~~
\frac{1}{\tilde \delta_-(x)}
\approx \ln\frac{1}{x},
\label{eq:nlo3}
\end{equation}
where
$\rho_g(x,\mu)$ are given by (\ref{rho.a}).

Because the ratio $f_g^-(x,Q^2)/f_g^+(x,Q^2)$ is rather small at the $Q^2$
values considered, the expression~(\ref{eq:pm3}) is modified to become
\begin{equation}
\hat{F}_k^Q(x,Q^2)\approx\tilde{M}_{k,g}(1,\mu^2,a)xf_g(x,\mu^2),
\end{equation}
where $\tilde{M}_{k,g}(1,\mu^2)$ is obtained from $M_{k,g}(n,\mu^2)$ by taking
the limit $n\to 1$ and replacing $1/(n-1)\to1/\tilde \delta_+$ in $B_{k,g}^{(1)}(n,a)$.
Consequently, one needs to substitute only
\begin{equation}
  B_{k,g}^{(1)}(1,a)\to\tilde{B}_{k,g}^{(1)}(1,a)
\end{equation}
in the NLO part of~(\ref{eq:exp}), i.e.
\be
\tilde{M}_{k,g}(1,Q^2,\mu^2) = e_Q^2a_s(\mu^2) \, \bigl[B^{(0)}_{k,g}(1,a) + a_s(\mu^2) \, \tilde{B}^{(1)}_{k,g}(1,a)\bigr].
\label{eq:expt}
\ee

\noindent
Using the identity
\begin{equation}
\frac{1}{I_0(\sigma(\hat{x}))}
\int^1_{\hat{x}}\frac{dy}{y}\beta\left(\frac{x}{y}\right) I_0(\sigma(y))
\approx \frac{1}{\tilde \delta_+(\hat{x})} + \phi_1(a)
\equiv \frac{1}{\hat{\delta}_+},~~
\hat{x}=\frac{x}{x_2} \, ,
\label{dp}
\end{equation}
where
\be
\phi_1(a)= -\ln (ax_2)-\hat{J}(a),~~ \hat{J}(a)=\frac{J(a)}{x_2} \, ,
\label{ph1}
\ee
we find the Mellin transform~(\ref{eq:mel}) of~(\ref{eq:nlo}) to be
\begin{equation}
\tilde{B}_{k,g}^{(1)}(1,a)\approx  \frac{1}{\hat{\delta}_+} \,
\bigl[R_{k,g}^{(1)}(1,a) + 4C_a B_{k,g}^{(0)}(1,a) L_{\mu}
  \bigr] \, ,
\label{eq:nloAa}
\end{equation}
with $R_{k,g}^{(1)}(1,a)$ and $B_{k,g}^{(0)}(1,a)$ are given in (\ref{eq:nloA}) and (\ref{Bk0}), respectively.

So, at the NLO we have for the ratio $\hat{R}^Q(x,Q^2)$:
\begin{equation}
  \hat{R}^Q_{\rm NLO}(x,Q^2)
  \approx\
  \frac{\tilde{M}_{L,g}(1,Q^2,m_f^2)}{\tilde{M}_{2,g}(1,Q^2,m_f^2)}
  = \frac{B_{L,g}^{(0)}(1,a) + a_s(\mu^2) \, \tilde{B}_{L,g}^{(1)}(1,a)
  }{B_{2,g}^{(0)}(1,a) + 
    a_s(\mu^2) \, \tilde{B}_{2,g}^{(1)}(1,a)
  }
  \, ,
\label{eq:triA}
\end{equation}
where the ratio has some $x$-dependence coming from the corresponding $x$-dependence of $\hat{\delta}_+$ in (\ref{dp}). The $x$-dependence
is in rather good agreement with the numerical results in (\ref{eq:tri}).

\subsection*{NNLO} \indent

At
NNLO, the coefficient function $C^g_{k}(x)$ has the form (\ref{6aNNLO}) with the NNLO coefficients $B^{(2)}_{k,g}(x,a)$ given in Eqs.
(\ref{eq:nnlo}) and (\ref{eq:nnloA}). Its moment
$M_{2,g}(n,Q^2,\mu^2)$ exhibits the corresponding structure
\be
M_{k,g}(n,Q^2,\mu^2) = e_Q^2a_s(\mu^2) \, \bigl[B^{(0)}_{k,g}(n,a) + a_s(\mu^2) \, B^{(1)}_{k,g}(n,a)+ a^2_s(\mu^2) \,
  B^{(2)}_{k,g}(n,a)\bigr] \, .
\label{eq:exp}
\ee

\noindent
The Mellin transforms of $B_{k,g}^{(2)}(x,a)$
exhibit singularities in the limit $n \to 1$,
which has the form
\be
\frac{1}{(n-1)^2} \to \frac{1}{\tilde{\delta}^2_{++}} = \frac{1}{\rho^2(x,\mu)}\, \frac{I_2(\sigma)}{I_0(\sigma)} \, ,
\label{sigmaPP}
\ee
where all definitions can be found in (\ref{4}) and (\ref{intro:1a}). 
So, by analogy with the NLO case, we have
\be
\tilde{M}_{k,g}(1,Q^2,\mu^2) = e_Q^2a_s(\mu^2) \, \bigl[B^{(0)}_{k,g}(1,a) + a_s(\mu^2) \, \tilde{B}^{(1)}_{k,g}(1,a)+ a^2_s(\mu^2) \,
  \tilde{B}^{(2)}_{k,g}(1,a)\bigr] \, .
\label{eq:exptN}
\ee
\noindent
Using the identity
\begin{equation}
\frac{1}{I_0(\sigma(\hat{x}))}
\int^1_{\hat{x}}\frac{dy}{y}\beta\left(\frac{x}{y}\right) \, \ln\left(\frac{y}{x}\right)
I_0(\sigma(y))
\approx \frac{1}{\tilde \delta^2_{++}(\hat{x})}
\equiv \frac{1}{\hat{\delta}^2_{++}},~~
\hat{x}=\frac{x}{x_1} \, ,
\label{dp2}
\end{equation}
we find the Mellin transform~(\ref{eq:mel}) of~(\ref{eq:nlo}) to be
\begin{equation}
\tilde{B}_{k,g}^{(2)}(1,a)\approx  \frac{1}{\hat{\delta}^2_{++}} \,
\bigl[R_{k,g}^{(2)}(1,a) + 4C_A R_{k,g}^{(1)}(1,a) L_{\mu} + 8C_A^2 B_{k,g}^{(0)}(1,a) L^2_{\mu}
  \bigr] 
\label{eq:nnloAa}
\end{equation}
with $R_{k,g}^{(2)}(1,a)$, $R_{k,g}^{(1)}(1,a)$ and $B_{k,g}^{(0)}(1,a)$ are given in  (\ref{eq:nnloA}), (\ref{eq:nloA}) and (\ref{Bk0}), respectively.


So,
  we have for the ratio $\hat{R}^Q(x,Q^2)$:
\begin{equation}
  \hat{R}^Q_{\rm NNLO}(x,Q^2)
  \approx\
  \frac{B_{L,g}^{(0)}(1,a) + \, a_s(\mu^2) \, \tilde{B}_{L,g}^{(1)}(1,a)+ \, a^2_s(\mu^2) \, \tilde{B}_{L,g}^{(2)}(1,a)}{
    B_{2,g}^{(0)}(1,a) + \, a_s(\mu^2) \, \tilde{B}_{2,g}^{(1)}(1,a) + \, a^2_s(\mu^2) \, \tilde{B}_{2,g}^{(2)}(1,a)}
  \, ,
\label{eq:triN}
\end{equation}
where the ratio has some $x$-dependence coming from the corresponding $x$-dependence of $\hat{\delta}_{++}$ in (\ref{dp2}). The $x$-dependence
is in rather good agreement with the numerical results in (\ref{eq:tri}).


\newpage

\begin{figure}
\begin{center}
\includegraphics[width=15cm]{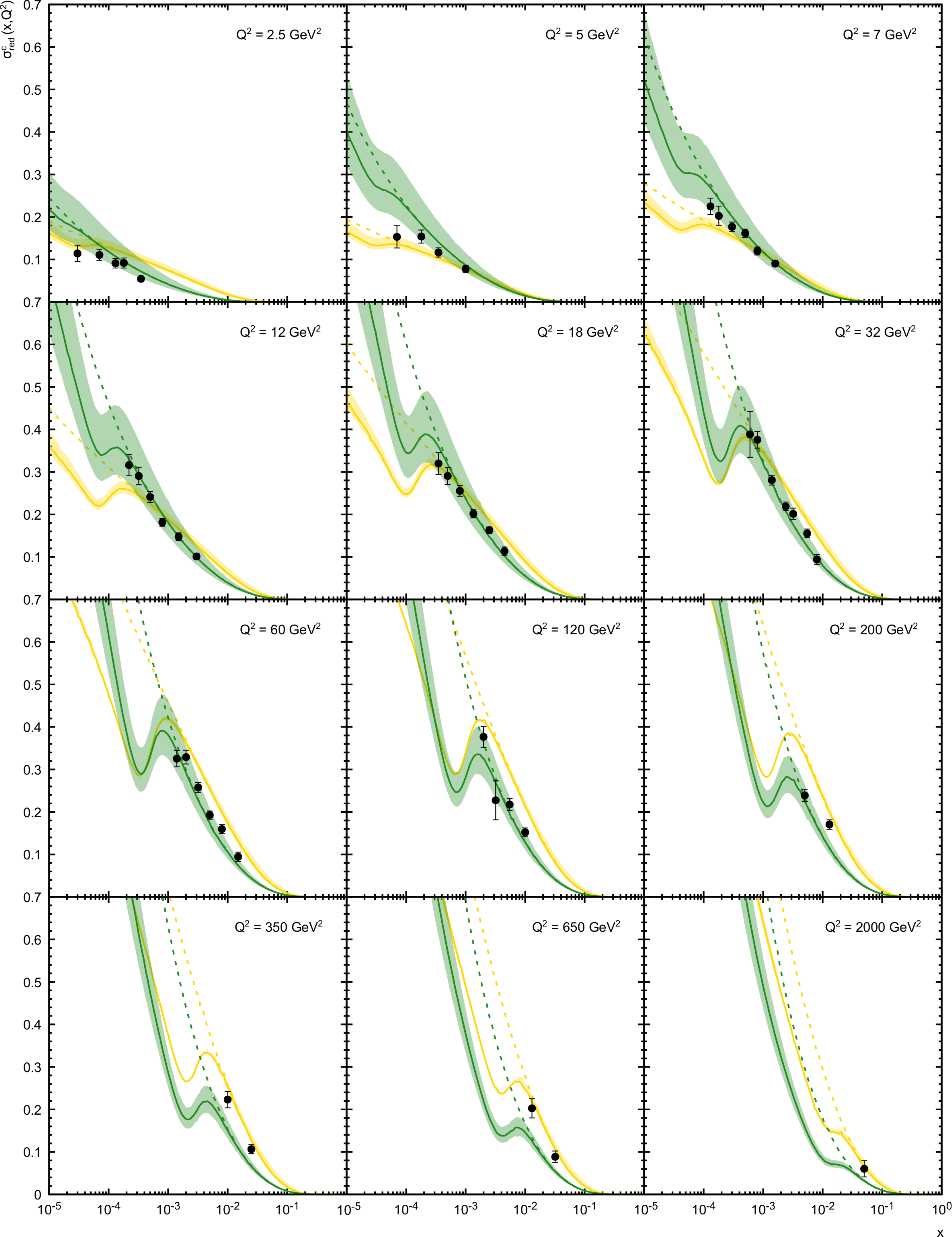}
\caption{The reduced charm cross sections $\sigma^{c\bar c}_{\rm red}(x,Q^2)$
  as a function of $x$ calculated at different $Q^2$ values. The predictions obtained with analytical 
  TMD gluon density in a proton and CCFM-evolved one are shown by the solid green and yellow 
  curves, respectively. The shaded bands correspond to the scale uncertainties of our calculations.
 The dashed curves represent the contributions from SF $F_2^c(x,Q^2)$, as it is described in the text.
  The experimental data are from H1 and ZEUS\cite{H1+ZEUS:2018}.
  }
\label{fig1}
\end{center}
\end{figure}

\begin{figure}
\begin{center}
\includegraphics[width=15cm]{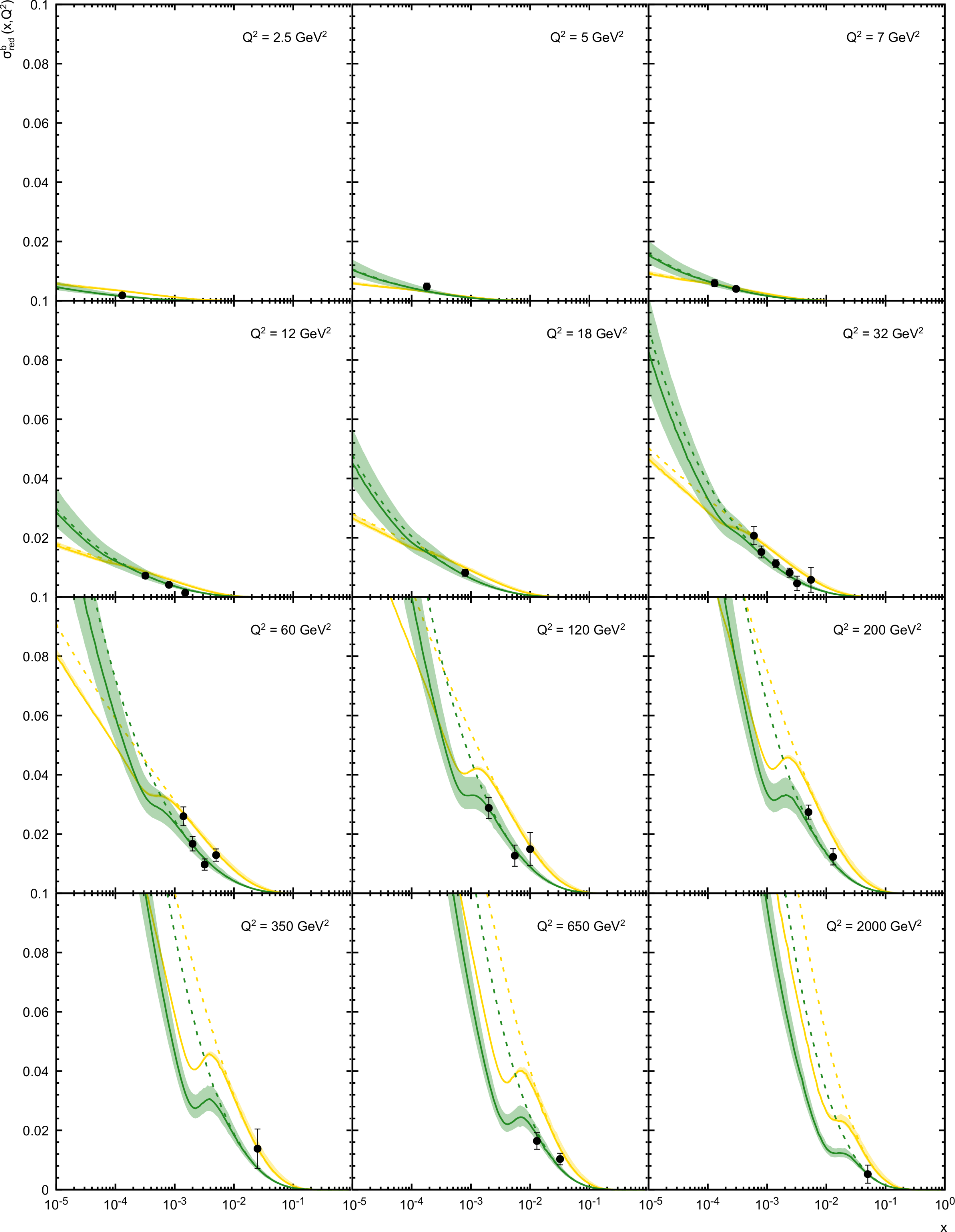}
\caption{
The reduced beauty cross sections $\sigma^{b\bar b}_{\rm red}(x,Q^2)$
  as a function of $x$ calculated at different $Q^2$ values. 
  Notation of all curves is the same as in Fig.~1.
  The experimental data are from H1 and ZEUS\cite{H1+ZEUS:2018}.
  }
\label{fig2}
\end{center}
\end{figure}


\begin{figure}
\begin{center}
\includegraphics[width=15cm]{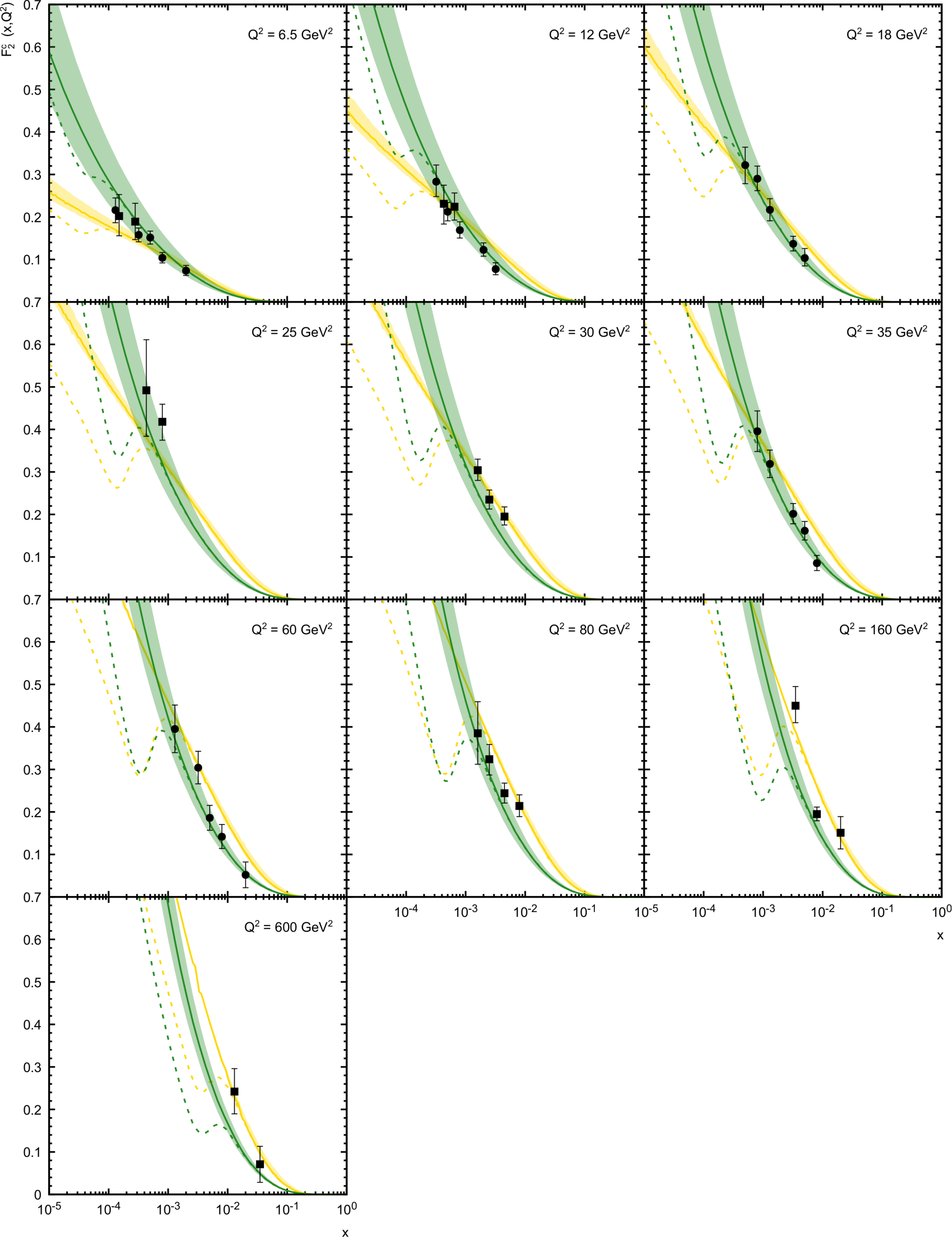}
\caption{The charm contribution to the proton structure function $F_2(x, Q^2)$
  as a function of $x$ calculated at different $Q^2$ values.
  Notation of all curves is the same as in Fig.~1.
The experimental data are from ZEUS\cite{69} and H1\cite{70}.}
\label{fig8}
\end{center}
\end{figure}

\begin{figure}
\begin{center}
\includegraphics[width=15cm]{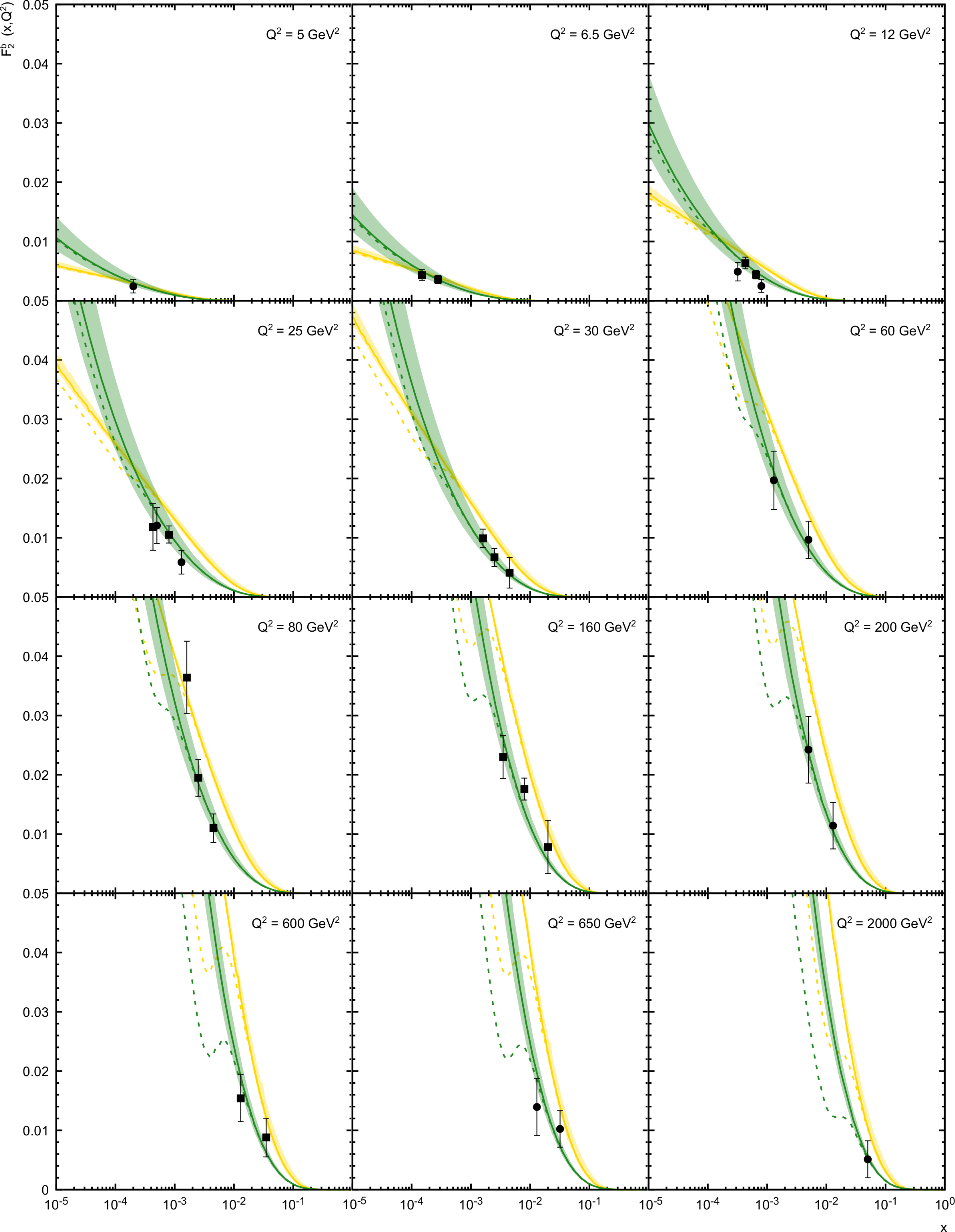}
\caption{The beauty contribution to the proton structure function $F_2(x, Q^2)$
  as a function of $x$ calculated at different $Q^2$ values.
  Notation of all curves is the same as in Fig.~1.
The experimental data are from ZEUS\cite{69} and H1\cite{71}.}
\label{fig9}
\end{center}
\end{figure}

\begin{figure}
\begin{center}
\includegraphics[width=15cm]{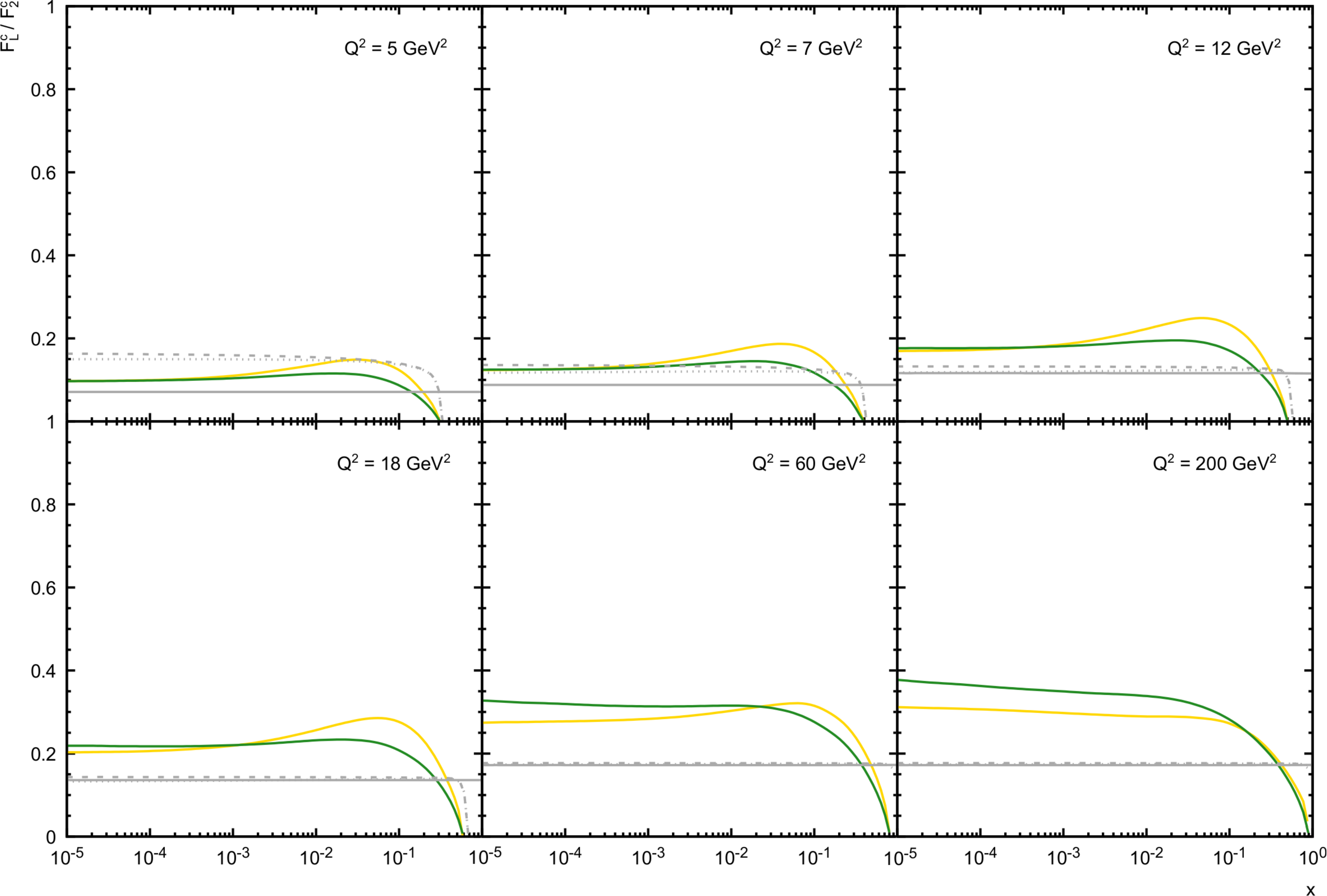}
\caption{The charm and beauty contribution to the ratio $\hat{R}^c(x, Q^2)$
  as a function of $x$ calculated at
   different $Q^2$ values. Notation of green and yellow solid curves is the same as in Fig.~1.
  The collinear results for $\hat{R}^c_{\rm LO}(x, Q^2)$, $\hat{R}^c_{\rm NLO}(x, Q^2)$
  and $\hat{R}^c_{\rm NNLO}(x, Q^2)$ are represented by solid, dashed and dotted gray curves, respectively.
}
\label{fig9}
\end{center}
\end{figure}

\end{document}